\documentclass[final,5p,times,twocolumn]{elsarticle}

\usepackage{graphicx}

\usepackage{amssymb}

\usepackage{lineno}

\journal{Elsevier}

\usepackage{multirow}
\usepackage[dvipdfmx]{pdflscape}
\usepackage{booktabs}
\usepackage{amsmath}

\newcommand\Pnt[2]{\mbox{$\mathrm{#1}_{#2}$}}
\newcommand\Tmax{\mbox{$\theta_\mathrm{max}$}}
\newcommand\Tcross{\mbox{$\theta_\mathrm{cross}$}}
\newcommand\arcdeg{\mbox{$^{\circ}$}}
\newcommand\Bezier{B\'{e}zier}
\newcommand\effmax{\mbox{$\epsilon\ (\theta<\Tmax)$}}
\newcommand\effcross{\mbox{$\epsilon\ (\theta<\Tcross)$}}
\newcommand\effmaxbg{\mbox{$\epsilon\ (\Tmax < \theta < 1.5\times\Tmax)$}}
\newcommand\effcrossbg{\mbox{$\epsilon\ (\Tcross < \theta < 1.5\times\Tmax)$}}

\begin{document}

\begin{frontmatter}



\title{Optimization of the Collection Efficiency of a Hexagonal Light Collector using\\ Quadratic and Cubic \Bezier\ Curves}


\author{Akira Okumura}\ead{oxon@mac.com}
\address{Solar-Terrestrial Environment Laboratory, Nagoya University, Furo-cho, Chikusa-ku, Nagoya, Aichi 464-8601, Japan}

\begin{abstract}

Reflective light collectors with hexagonal entrance and exit apertures are frequently used in front of the focal-plane camera of a very-high-energy gamma-ray telescope to increase the collection efficiency of atmospheric Cherenkov photons and reduce the night-sky background entering at large incident angles. The shape of a hexagonal light collector is usually based on Winston's design, which is optimized for only two-dimensional optical systems. However, it is not known whether a hexagonal Winston cone is optimal for the real three-dimensional optical systems of gamma-ray telescopes. For the first time we optimize the shape of a hexagonal light collector using quadratic and cubic \Bezier\ curves. We demonstrate that our optimized designs simultaneously achieve a higher collection efficiency and background reduction rate than traditional designs.

\end{abstract}


\end{frontmatter}


\section{Introduction}
\label{sec_introduction}

\begin{figure}[!tb]
  \begin{center}
    \includegraphics[width=7cm]{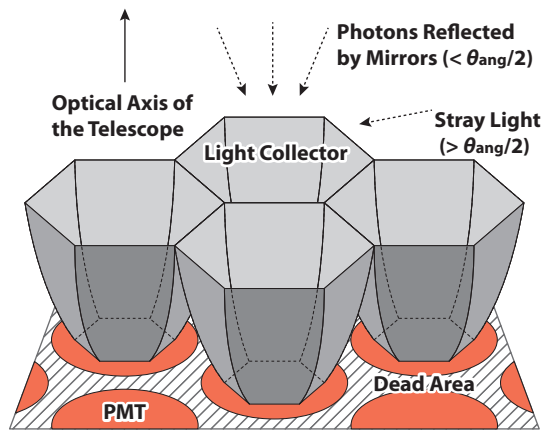}
    \caption{Schematic of a typical focal-plane camera of an IACT. Circles show the effective area of PMTs aligned in a honeycomb structure. Hatched region is a dead area not covered by the photocathodes of the PMTs. Hexagonal light collectors are placed in front of the photocathodes. Cherenkov photons enter the light collectors from the direction of the mirrors within the numerical aperture of the telescope optical system, whereas stray light enters at large incident angles.}
    \label{fig_FP}
  \end{center}
\end{figure}

The photon detection efficiency of a focal-plane camera is a significant determinants of the gamma-ray detection sensitivity of an imaging atmospheric Cherenkov telescope (IACT). To increase the signal-to-noise ratio of faint and transient ($\sim5$~ns) Cherenkov photons induced by gamma-ray air showers against the dominant night-sky background, a substantial amount of effort has been dedicated for developing faster waveform sampling systems ($>500$~MHz), multiple telescopes with larger apertures ($>10$~m), photodetectors with higher quantum efficiency ($>25$\%), and mirrors with higher reflectivity ($>95$\%). Design studies of light collectors have also been conducted to guide Cherenkov photons onto the effective area of photodetectors such as photomultiplier tubes (PMTs)  at higher collection efficiencies. This is because a considerable area of the focal plane of an IACT is not covered by photodetectors when they are aligned in a honeycomb structure as shown in Figure~\ref{fig_FP}. Hence some of photons focused on the focal plane can not be detected. Using hexagonal light collectors consisting of reflective surfaces is an option for reducing this dead area.

Light collectors have been widely used in various gamma-ray and cosmic-ray telescopes, and design studies on the shape of the light collectors have been conducted together \cite{Baltrusaitis:1985:The-Utah-Flys-Eye-Detector,Kabuki:2003:Development-of-an-atmospheric-Cherenkov-,Bernlohr:2003:The-optical-system-of-the-H.E.S.S.-imagi,Radu:2000:Design-studies-for-nonimaging-light-conc,Pare:2002:CELESTE:-an-atmospheric-Cherenkov-telesc}. The first requirement of such a light collector is to gather maximum photons reflected by the telescope mirrors. The second is to minimize the collection efficiency of stray light with incident angles larger than half of the angular aperture ($\theta_\mathrm{ang}/2$) of the optical system, because the night-sky background can enter the focal-plane camera from the night sky directly or from the ground indirectly (Figure~\ref{fig_FP}). A well-designed light collector for an IACT must satisfy these requirements simultaneously in order to achieve a higher signal-to-noise ratio.

\begin{figure*}[!tb]
  \begin{center}
    \includegraphics[width=15cm]{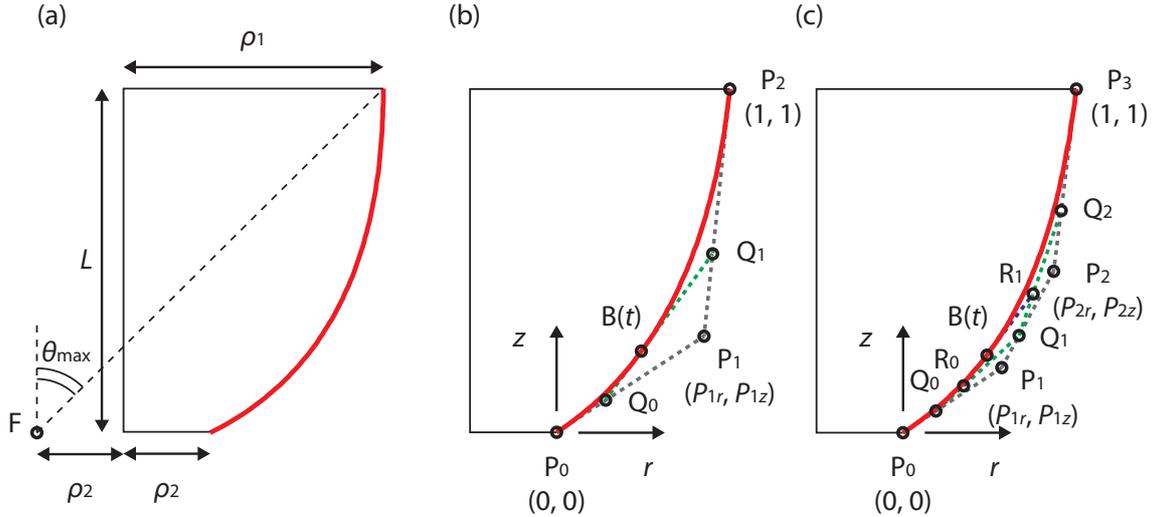}
    \caption{(a) Vertical section of a Winston cone. Thick red curve is the part of a parabolic curve whose focus is located at point F. Vertical solid line is the optical axis of the collector. (b) Vertical section of a quadratic \Bezier\ light collector. Thick red curve is the track of a parameterized point $\mathrm{B}(t)$ defined by three control points: $\Pnt{P}{0}$, $\Pnt{P}{1}$, and $\Pnt{P}{2}$, where $\Pnt{P}{0}\Pnt{Q}{0}:\Pnt{Q}{0}\Pnt{P}{1}=\Pnt{P}{1}\Pnt{Q}{1}:\Pnt{Q}{1}\Pnt{P}{2}=\Pnt{Q}{0}\mathrm{B}(t):\mathrm{B}(t)\Pnt{Q}{1}=t:(1-t) \quad (0\le t \le 1)$.  The coordinates of point $\Pnt{P}{1}$ are written in relative values between $\Pnt{P}{0}(r=0, z=0)$ and $\Pnt{P}{2}(r=1, z=1)$. (c) Vertical section of a cubic \Bezier\ light collector. Thick red curve is the track of a parameterized point $\mathrm{B}(t)$ defined by four control points: $\Pnt{P}{0}$, $\Pnt{P}{1}$, $\Pnt{P}{2}$, and $\Pnt{P}{3}$, where $\Pnt{P}{0}\Pnt{Q}{0}:\Pnt{Q}{0}\Pnt{P}{1}=\Pnt{P}{1}\Pnt{Q}{1}:\Pnt{Q}{1}\Pnt{P}{2}=\Pnt{P}{2}\Pnt{Q}{2}:\Pnt{Q}{2}\Pnt{P}{3}=\Pnt{Q}{0}\Pnt{R}{0}:\Pnt{R}{0}\Pnt{Q}{1}=\Pnt{Q}{1}:\Pnt{R}{1}\Pnt{Q}{2}=\Pnt{R}{0}\mathrm{B}(t):\mathrm{B}(t)\Pnt{R}{1}=t:(1-t) \quad (0\le t \le 1)$.}
    \label{fig_Bezier}
  \end{center}
\end{figure*}

\begin{figure}[!b]
  \begin{center}
    \includegraphics[width=8cm]{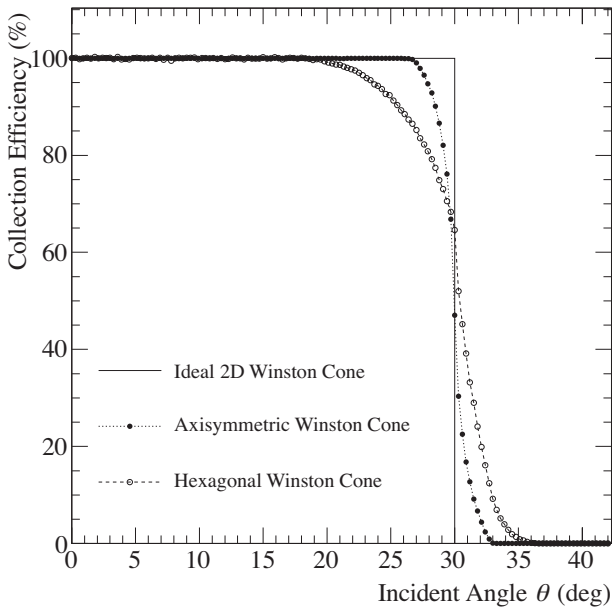}
    \caption{Comparison of simulated collection efficiencies of two-dimensional (solid line), axisymmetric (dotted with filled circles), and hexagonal (dashed with open circles) Winston cones. We assumed $\rho_1$, $\rho_2$, and $R$ to be $20$~mm, $10$~mm, and $1.0$, respectively. Note that the data points for the axisymmetric cone must be scaled by $\pi/2\sqrt{3}\fallingdotseq0.907$ when the cones are aligned in a honeycomb structure, because the circular entrance aperture has a dead area. The typical statistical error of the graphs is $0.3$\%.}
    \label{fig_Winston}
  \end{center}
\end{figure}

The basic design concept of a light collector for the two-dimensional case was first developed by \citet{Winston:1970:Light-Collection-within-the-Fr} using two inclined paraboloids. Figure~\ref{fig_Bezier}(a) shows a schematic of a Winston light collector (hereafter a Winston cone). The cutoff angle \Tmax\ and the height $L$ of the cone are uniquely defined as $\Tmax=\arcsin(\rho_2/\rho_1)$ and $L = (\rho_1 + \rho_2)/\tan\Tmax$, where $\rho_1$ and $\rho_2$ are the radii of the entrance and exit apertures, respectively. Ideally, this design can gather $100$\% of the photons for incident angles less than \Tmax, and $0$\% for larger incident angles, assuming the reflectivity of the light collector $R$ to be 100\%. In addition, Winston demonstrated that an axisymmetric cone using the same paraboloid had an excellent collection efficiency, as shown in Figure~\ref{fig_Winston}.

In reality, neither the ideal two-dimensional nor the axisymmetric Winston cone is used for the focal-plane cameras in IACTs. Instead, three-dimensional light collectors with hexagonal entrance and exit apertures are used because they can cover the entire focal plane, as illustrated in Figure~\ref{fig_FP}. In this situation, the shape of the six side surfaces of a light collector is usually given by Winston formulation \cite{Kabuki:2003:Development-of-an-atmospheric-Cherenkov-,Bernlohr:2003:The-optical-system-of-the-H.E.S.S.-imagi}, although it is not optimized for a hexagonal cone.

It is not known whether Winston design is optimal for the side surfaces of a hexagonal light collector because the paraboloid optimized for the two-dimensional space cannot collect some of the skew rays onto the exit apertures in three-dimensional space. Figure~\ref{fig_Winston} compares the collection efficiencies of two-dimensional, axisymmetric, and hexagonal Winston cones. The two-dimensional cone has an ideal discontinuous cutoff at $\Tmax=30\arcdeg$, whereas the axisymmetric cone has a continuous cutoff, and the hexagonal cone has a more gradual cutoff around \Tmax\ because of the contribution from skew rays.

In this study, we demonstrate that a hexagonal light collector, which has a better collection efficiency than the normal hexagonal Winston cone, can be designed using quadratic or cubic \Bezier\ curves instead of Winston's original paraboloid. The parameters for the optimized designs found in our ray-tracing simulations are given in Table~\ref{tab_coords} so that readers can use them for their own applications.

\section{Method}
\subsection{\Bezier\ Curve}

To search a hexagonal light collector having the maximum collection efficiency, we tweaked the shape of the six side surfaces using quadratic or cubic \Bezier\ curves. A \Bezier\ curve is a smooth parametric curve often used in computer graphics and computer-aided design \cite{Bezier:1978:General-distortion-of-an-ensemble-of-bip}. The coordinates of the curve are given by a single parameter $t\ (0\le t \le 1)$ and two or more control points $\Pnt{P}{i}\ (i = 0, 1 \dots N)$, where $N$ is the order of the curve.

For $N=2$, a quadratic \Bezier\ curve is given by
\begin{equation}
\vec{B}(t) = (1-t)^2\vec{P}_0 + 2(1-t)t\vec{P}_1 + t^2\vec{P}_2,
\end{equation}
where, as shown in Figure~\ref{fig_Bezier}(b), \Pnt{P}{0} and \Pnt{P}{2} are located at the end points of the entrance and exit apertures, respectively. Various \Bezier\ curves can be generated by changing the coordinates of control point \Pnt{P}{1} ($P_{1r}$ and $P_{1z}$).

For $N=3$, a cubic \Bezier\ curve (Figure~\ref{fig_Bezier}(c)) is similarly given as follows.
\begin{equation}
\vec{B}(t) = (1-t)^3\vec{P}_0 + 3(1-t)^2t\vec{P}_1 + 3(1-t)t^2\vec{P}_2 + t^3\vec{P}_3.
\end{equation}
Additional free coordinates, $P_{2r}$ and $P_{2z}$, enable us to generate curves more flexibly.

\subsection{Ray-tracing Simulator}

\begin{figure}[!tb]
  \begin{center}
    \includegraphics[width=9cm]{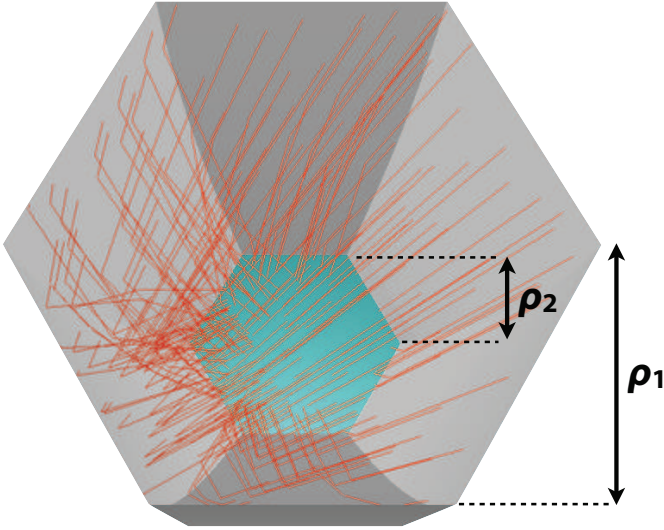}
    \caption{Example of a hexagonal light collector simulated with the ROBAST library. Red segments are the tracks of simulated photons. Hexagonal blue surface is the input window of a PMT. The geometry of the light collector was built with the \texttt{AGeoBezierPgon} class in ROBAST.}
    \label{fig_ROBAST}
  \end{center}
\end{figure}

We used a ray-tracing simulator, ROOT-based simulator for ray tracing (ROBAST), for our light collector simulations \cite{Okumura:2011:Development-of-Non-sequential-Ray-tracin}. The non-sequential photon-tracking engine provided with ROBAST and ROOT geometry library \cite{Brun:2003:The-ROOT-geometry-package} is essential for our study because incident photons can be reflected multiple times on the surfaces of a light collector. In addition, it is easy for the user to add a new geometry class to the ROBAST library; hence, we can flexibly simulate optical components of various shapes. Figure~\ref{fig_ROBAST} shows an example of a hexagonal light collector whose side surfaces are defined by a cubic \Bezier\ curve.

\subsection{Simulations}

\begin{figure}
  \begin{center}
    \includegraphics[width=9cm]{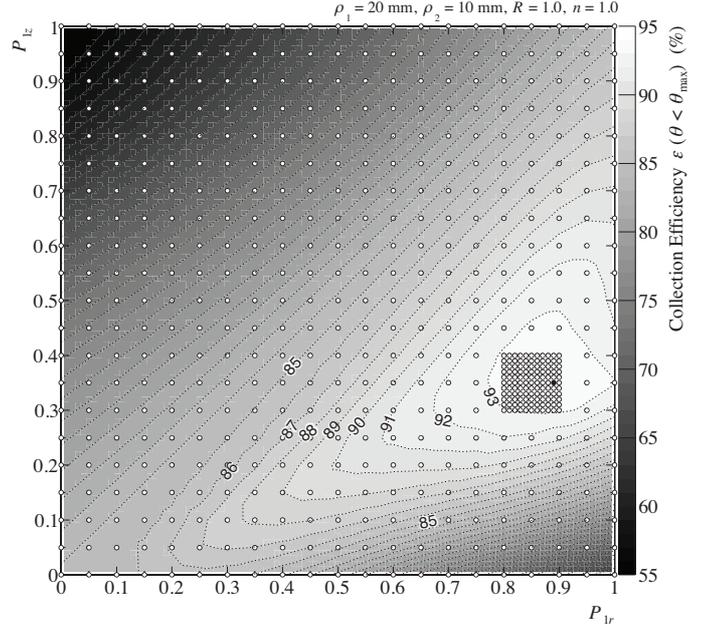}
    \caption{Contour map of the collection efficiency \effmax\ of a hexagonal light collector defined by a quadratic \Bezier\ curve having $\rho_1$, $R$, and $n$ values of $20$~mm, $1.0$, and $1.0$, respectively. Open circles show the scanned coordinates of control point \Pnt{P}{1}. One ray-tracing simulation was done for each coordinate pair. We first scanned the coordinates in $0.05$ steps before scanning at a finer mesh of $0.01$. Black filled circle indicates the coordinates that maximize \effmax.}
    \label{fig_canContour}
  \end{center}
\end{figure}

\begin{figure*}[!t]
  \begin{center}
    \includegraphics[width=18cm]{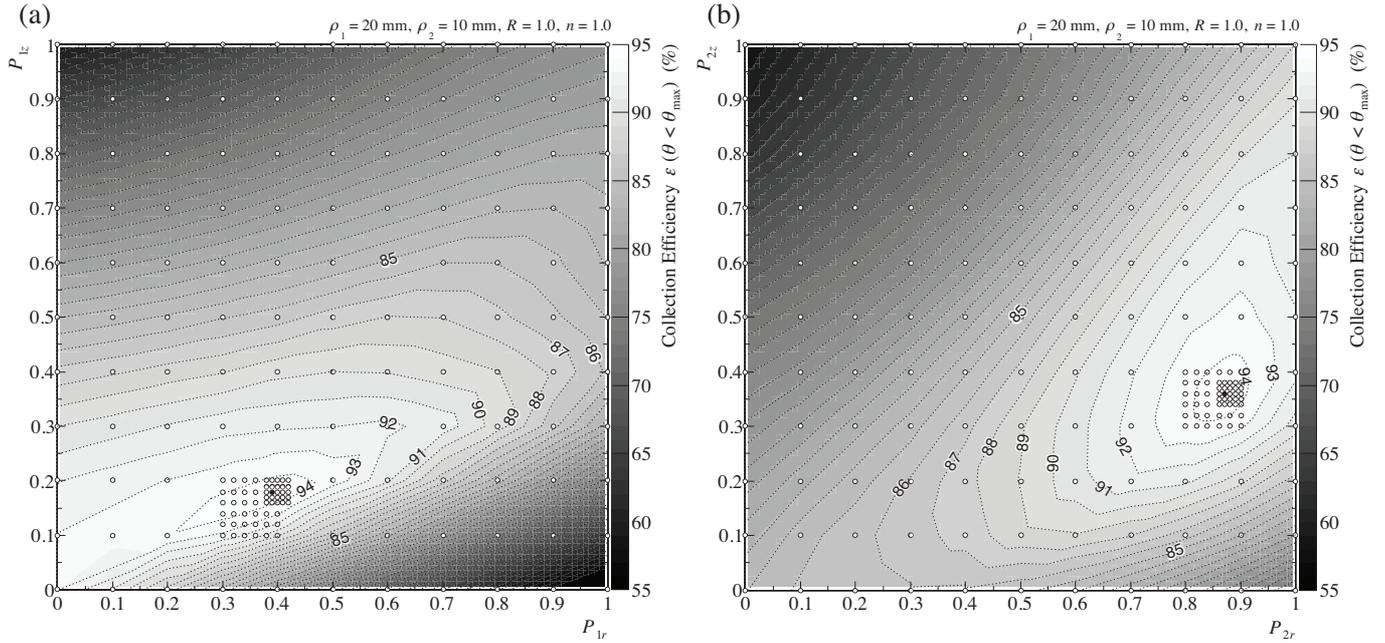}
    \caption{Contour maps of the collection efficiency \effmax\ of a hexagonal light collector defined by a cubic \Bezier\ curve ($\rho_1=20$~mm, $R=1.0$, $n=1.0$). Symbols are defined as in Figure~\ref{fig_canContour}. The intervals between coordinates were changed in the order of $0.1$, $0.02$, and $0.01$. (a) $P_\mathrm{1r}$--$P_\mathrm{1z}$ space. (b) $P_\mathrm{2r}$--$P_\mathrm{2z}$ space.}
    \label{fig_canContourCubic}
  \end{center}
\end{figure*}

We simulated various types of hexagonal light collectors by changing three sets of parameters. The first parameter set is the coordinates of the movable control points: \Pnt{P}{1} for quadratic \Bezier\ curves, and \Pnt{P}{1} and \Pnt{P}{2} for cubic \Bezier\ curves. The relative coordinates of the control points are defined as shown in Figure~\ref{fig_Bezier}(b) and (c). They were scanned from $0$ to $1$ in steps of at least $0.01$. Figure~\ref{fig_canContour} shows a contour map of the collection efficiency \effmax\ of a quadratic-\Bezier-type light collector having $\rho_1$ and $\rho_2$ values of $20$~mm and $10$~mm, respectively. Here,\effmax\ is the angular-weighted collection efficiency integrated over $\theta$ ($0 \le \theta < \Tmax$). Hereafter, we maximize \effmax\ in our simulations. As shown in the figure, \effmax\ is maximum when $P_\mathrm{1r}$ and $P_\mathrm{1z}$ are set to $0.89$ and $0.35$, respectively.

The second parameter set consists of the entrance aperture $\rho_1$ and the exit aperture $\rho_2$. We scanned $\rho_1$ from $18$~mm ($\Tmax=33.7^\circ$) to $30$~mm ($\Tmax=19.5^\circ$) in steps of $1$~mm in order to cover the typical range of opening angles in the optical systems of gamma-ray telescopes.\footnote{$\theta_\mathrm{ang}/2$ is $31.0^\circ$ for an optical system of $f/D=1.2$, and $26.6^\circ$ for $f/D=1.0$, where $f$ and $D$ are the focal length and the mirror diameter of the optical system, respectively.} In contrast, we fixed $\rho_2$ at $10$~mm because the optical performance of an optimized light collector is determined by the ratio of $\rho_1$ to $\rho_2$. We use the same definition for the cone height $L$ and the cutoff angle \Tmax\ as those used for Winston cones in order to reduce the number of free parameters.

The third set consists of the reflectivity $R$ of the light collector and the refractive index $n$ of the input window of a PMT. For an ideal case, they were assumed to be $1.0$ and $1.0$, respectively. For a more realistic case, $R=0.9$ and $n=1.5$ were used. In the latter case, $10$\% of the reflected photons are randomly absorbed by the cone surfaces, and angular-dependent Fresnel reflection ($4$\% for normal incident photons) is considered at the boundary between the air and the input window of a PMT. We assumed that $100$\% of the photons that propagated to the boundary between the input window and the photocathode were detected. The waveform dependence of the reflectivity and the quantum efficiency of the photocathode were not considered.

For each parameter set, the incident angle $\theta$ was changed from $0$ to $1.5\times\Tmax$ in steps of $\Tmax/100$, and the polar angle $\phi$ was changed from $0^\circ$ to $59.7^\circ$ in steps of $0.3^\circ$. We traced $500$ photons randomly scattered on the entrance aperture for each pair of $\theta$ and $\phi$, and averaged the collection efficiency over the polar angles.

\section{Results}

\begin{figure*}
  \begin{center}
    \includegraphics[width=18cm]{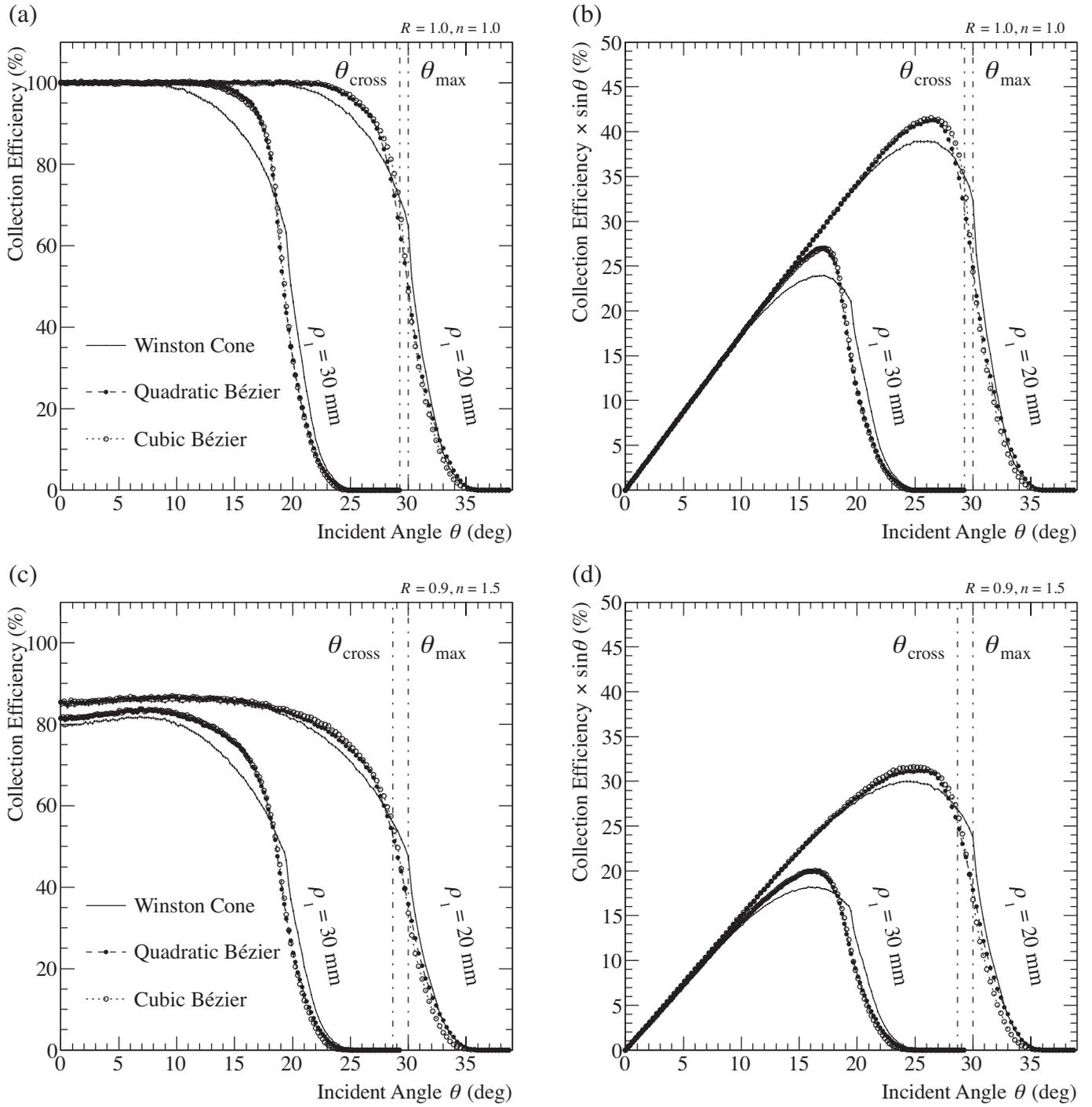}
    \caption{(a) Incident angle dependency of the collection efficiencies of hexagonal light collectors ($\rho_1=20$~mm and $\rho_1=30$~mm, and $\rho_2=10$~mm). Solid lines are the collection efficiencies of hexagonal Winston cones without any optimization. Filled and open circles represent optimized quadratic and cubic \Bezier\ cones, respectively. Values of \Tmax\ and \Tcross\ for the $\rho_1=20$~mm case are marked with dot-dashed lines. (b) Same as (a), but the vertical axis is angularly weighted by $\sin\theta$. (c) Same as (a), but $R$ and $n$ are $0.9$ and $1.5$, respectively. (d) Same as (b), but $R$ and $n$ are $0.9$ and $1.5$, respectively.}
    \label{fig_canCurve}
  \end{center}
\end{figure*}

We obtained the optimal coordinates of the control point(s) of quadratic and cubic \Bezier\ curves that maximize the angular-averaged collection efficiency \effmax\ for each combination of $\rho_1$, $R$, and $n$. The optimal coordinates for quadratic \Bezier\ curves were identified using contour plots of \effmax, as already shown in Figure~\ref{fig_canContour}. For a cubic \Bezier\ curve, $P_\mathrm{1r}$, $P_\mathrm{1z}$, $P_\mathrm{2r}$, and $P_\mathrm{2z}$ were scanned independently, as shown in Figure~\ref{fig_canContourCubic}, but only two combinations, $P_\mathrm{1r}$--$P_\mathrm{1z}$ and $P_\mathrm{2r}$--$P_\mathrm{2z}$ spaces, are shown. There is a single maximum at the optimum coordinates, and the efficiency varies smoothly. The optimized coordinates for all the combinations used in the simulations are tabulated in Table~\ref{tab_coords}.

\begin{table*}
\begin{center}
\caption{Relative coordinates of the control points for the optimized quadratic or cubic \Bezier\ curves for each set of $\rho_1$ and $\rho_2$.}
\label{tab_coords}
\begin{tabular}{cccccccccccccc}
\hline
\hline
&& \multicolumn{6}{c}{$R=1.0$, $n=1.0$} & \multicolumn{6}{c}{$R=0.9$, $n=1.5$} \\ \cmidrule(lr){3-8} \cmidrule(lr){9-14}
&& \multicolumn{2}{c}{Quad. \Bezier} & \multicolumn{4}{c}{Cubic \Bezier} & \multicolumn{2}{c}{Quad. \Bezier} & \multicolumn{4}{c}{Cubic \Bezier} \\ \cmidrule(lr){3-4} \cmidrule(lr){5-8} \cmidrule(lr){9-10} \cmidrule(lr){11-14}
$\rho_1$ & $\rho_2$ & $P_{1r}$ & $P_{1z}$ & $P_{1r}$ & $P_{1z}$ & $P_{2r}$ & $P_{2z}$ & $P_{1r}$ & $P_{1z}$ & $P_{1r}$ & $P_{1z}$ & $P_{2r}$ & $P_{2z}$ \\
(mm) & (mm) & & & & & & & & & & & & \\ \hline
$18$ & $10$ & $0.90$ & $0.38$ & $0.26$ & $0.14$ & $0.88$ & $0.33$ & $0.94$ & $0.36$ & $0.33$ & $0.17$ & $0.88$ & $0.34$ \\
$19$ & $10$ & $0.89$ & $0.36$ & $0.33$ & $0.16$ & $0.88$ & $0.34$ & $0.94$ & $0.36$ & $0.29$ & $0.15$ & $0.89$ & $0.33$ \\
$20$ & $10$ & $0.89$ & $0.35$ & $0.39$ & $0.18$ & $0.87$ & $0.36$ & $0.90$ & $0.35$ & $0.18$ & $0.11$ & $0.89$ & $0.30$ \\
$21$ & $10$ & $0.89$ & $0.35$ & $0.29$ & $0.14$ & $0.87$ & $0.32$ & $0.90$ & $0.34$ & $0.42$ & $0.19$ & $0.88$ & $0.37$ \\
$22$ & $10$ & $0.87$ & $0.33$ & $0.28$ & $0.13$ & $0.88$ & $0.31$ & $0.90$ & $0.34$ & $0.30$ & $0.14$ & $0.88$ & $0.32$ \\
$23$ & $10$ & $0.87$ & $0.33$ & $0.50$ & $0.20$ & $0.87$ & $0.41$ & $0.90$ & $0.33$ & $0.30$ & $0.14$ & $0.88$ & $0.30$ \\
$24$ & $10$ & $0.87$ & $0.32$ & $0.34$ & $0.14$ & $0.88$ & $0.34$ & $0.90$ & $0.32$ & $0.19$ & $0.10$ & $0.88$ & $0.27$ \\
$25$ & $10$ & $0.86$ & $0.31$ & $0.54$ & $0.20$ & $0.88$ & $0.42$ & $0.88$ & $0.32$ & $0.28$ & $0.12$ & $0.88$ & $0.29$ \\
$26$ & $10$ & $0.86$ & $0.31$ & $0.62$ & $0.22$ & $0.89$ & $0.50$ & $0.89$ & $0.31$ & $0.25$ & $0.11$ & $0.86$ & $0.27$ \\
$27$ & $10$ & $0.86$ & $0.30$ & $0.60$ & $0.21$ & $0.89$ & $0.48$ & $0.88$ & $0.30$ & $0.19$ & $0.09$ & $0.87$ & $0.25$ \\
$28$ & $10$ & $0.85$ & $0.29$ & $0.62$ & $0.21$ & $0.89$ & $0.50$ & $0.87$ & $0.30$ & $0.24$ & $0.10$ & $0.89$ & $0.26$ \\
$29$ & $10$ & $0.86$ & $0.28$ & $0.63$ & $0.21$ & $0.89$ & $0.51$ & $0.87$ & $0.29$ & $0.22$ & $0.09$ & $0.88$ & $0.25$ \\
$30$ & $10$ & $0.85$ & $0.27$ & $0.65$ & $0.21$ & $0.89$ & $0.53$ & $0.87$ & $0.28$ & $0.21$ & $0.09$ & $0.88$ & $0.25$ \\
\hline
\hline
\end{tabular}
\end{center}
\end{table*}

Figure~\ref{fig_canCurve}(a) shows the collection efficiency vs. incident angle $\theta$ for four optimized ideal light collectors: quadratic or cubic \Bezier\ curves, and $\rho_1=20$ or $30$~mm. In each case, the optimized light collector has a sharper cutoff than the traditional hexagonal Winston cone with the same $\rho_1$, $\rho_2$, and $L$. In addition, the collection efficiencies of the optimized cones are higher in a wider range of incident angles but a bit worse at $\Tmax$. The cubic \Bezier\ cone exhibits slightly better performance than the quadratic \Bezier\ cone. This is because the cubic curve is more flexible than the quadratic owing to the additional control point \Pnt{P}{2}.

In Figure~\ref{fig_canCurve}(b), we show the collection efficiency weighted by $\sin\theta$ in order to clarify the contribution by solid angle around \Tmax. The \effmax\ values of the Winston, the optimized quadratic \Bezier, and the optimized cubic \Bezier\ cones can be compared by integrating the graphs over $\theta$. For $\rho_1=20$~mm, \effmax\ is $93.1$\%, $93.7$\%, and $94.4$\% for the three cones, respectively. For $\rho_1=30$~mm, the values are $89.3$\%, $92.3$\%, and $92.6$\%, respectively. In each case, the optimized cones can achieve higher collection efficiencies by a few percent for signal photons with incident angles of less than \Tmax.

In addition to the higher collection efficiencies for the signal, \effmaxbg, the angular-averaged collection efficiencies between $\theta=\Tmax$ and $1.5\times\Tmax$, become much smaller for the optimized cones. For $\rho_1=20$~mm, the \effmaxbg\ values of the Winston, the optimized quadratic \Bezier, and the optimized cubic \Bezier\ cones are $5.22$\%, $4.62$\%, and $3.99$\%, respectively. For $\rho_1=30$~mm, these values are $8.15$\%, $5.53$\%, and $5.34$\%, respectively. This means that we can reduce the night-sky background entering with large incident angles by $\sim20-30$\%.

Here, we introduce a new parameter, \Tcross, at which the efficiency curves for the Winston and optimized cubic \Bezier\ cones cross each other, as shown in Figures~\ref{fig_canCurve}(a) and (b). This is because if we set \Tmax\ to $\theta_\mathrm{ang}/2$, then more than $5$\% of the signal photons are lost. Therefore, we should use a smaller angle as the cutoff angle of a light collector. We tentatively use \Tcross\ for the cutoff angle. We tabulate \effmax, \effmaxbg, \effcross, and \effcrossbg\ in Table~\ref{tab_eff1}. For example, for $\rho_1=20$~mm, $R=1$, and $n=1$, the \effcross\ values for the Winston and optimized cubic \Bezier\ cones are $94.6$\% and $96.7$\%, respectively. The \effcrossbg\ values are $8.13$\% and $6.48$\%, respectively. Therefore, using a cubic \Bezier\ curve, we can achieve a $2.2$\% higher collection efficiency for signal photons, and an $20.3$\% lower efficiency for stray background light.

\begin{figure}
  \begin{center}
    \includegraphics[width=9cm]{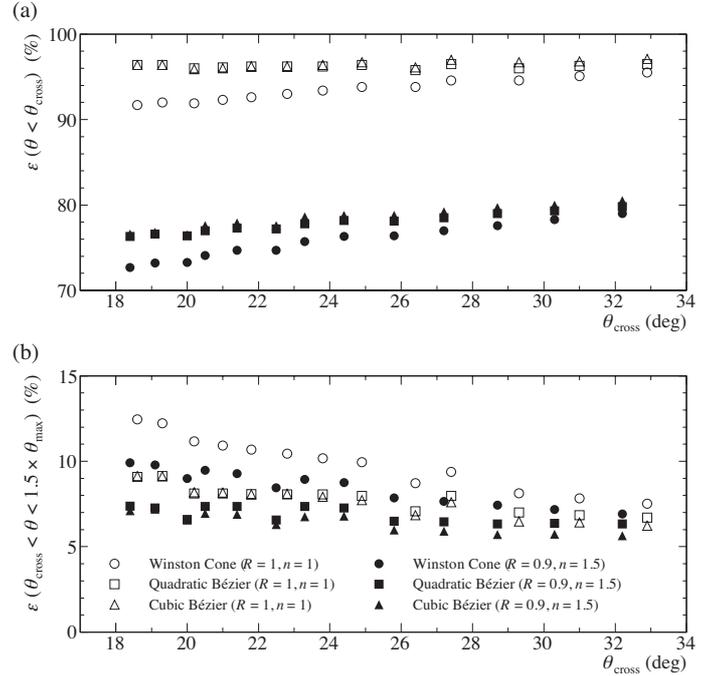}
    \caption{(a) \effcross\ vs. \Tcross. Circles, rectangles, and triangles represent \effcross\ for hexagonal Winston cones,.optimized quadratic \Bezier\ cones, and optimized cubic \Bezier\ cones, respectively. Open symbols indicate the ideal case in which $R=1.0$ and $n=1.0$. Filled symbols indicate the realistic case in which $R=0.9$ and $n=1.5$. (b) \effcrossbg\ vs. \Tcross.}
    \label{fig_canGain}
  \end{center}
\end{figure}

{\tabcolsep = 1.5mm
\begin{landscape}
\begin{table}
\caption{Simulated collection efficiencies of light collectors, assuming the reflectivity of the light collectors and the refractive index of the input windows of PMTs are $1.0$ and $1.0$, respectively. Positive and negative values in parentheses are the relative gains (\%) in collection efficiency compared to the regular Winston cones.}
\label{tab_eff1}
\begin{center}
\begin{tabular}{ccccccccccccccccccc}
\hline
\hline
&&&&& \multicolumn{3}{c}{\effmax} & \multicolumn{3}{c}{\effmaxbg} & \multicolumn{3}{c}{\effcross} & \multicolumn{3}{c}{\effcrossbg} \\ \cmidrule(lr){6-8} \cmidrule(lr){9-11} \cmidrule(lr){12-14} \cmidrule(lr){15-17}

\multicolumn{1}{c}{$\rho_1$} & \multicolumn{1}{c}{$\rho_2$} & \multicolumn{1}{c}{$L$} & \multicolumn{1}{c}{\Tmax} & \multicolumn{1}{c}{\Tcross} & Win. & Quad. & Cubic & Win. & Quad. & Cubic & Win. & Quad. & Cubic & Win. & Quad. & Cubic \\

\multicolumn{1}{c}{(mm)} & \multicolumn{1}{c}{(mm)} & \multicolumn{1}{c}{(mm)} & \multicolumn{1}{c}{($^\circ$)} & \multicolumn{1}{c}{($^\circ$)} & \multicolumn{1}{c}{(\%)} & \multicolumn{1}{c}{(\%)} & \multicolumn{1}{c}{(\%)} & \multicolumn{1}{c}{(\%)} & \multicolumn{1}{c}{(\%)} & \multicolumn{1}{c}{(\%)} & \multicolumn{1}{c}{(\%)} & \multicolumn{1}{c}{(\%)} & \multicolumn{1}{c}{(\%)} & \multicolumn{1}{c}{(\%)} & \multicolumn{1}{c}{(\%)} & \multicolumn{1}{c}{(\%)} \\ \hline
$18$ & $10$ & $41.9$ & $33.7$ & $32.9$ & $94.0$ & $94.2\ (+0.3)$ & $95.0\ (+1.1)$ & $4.50$ & $4.24\ (-5.8)$ & $3.53\ (-21.5)$ & $95.5$ & $96.5\ (+1.0)$ & $97.1\ (+1.6)$ & $7.51$ & $6.70\ (-10.9)$ & $6.22\ (-17.2)$ \\
$19$ & $10$ & $46.9$ & $31.8$ & $31.0$ & $93.5$ & $94.0\ (+0.4)$ & $94.7\ (+1.2)$ & $4.87$ & $4.44\ (-8.8)$ & $3.77\ (-22.6)$ & $95.1$ & $96.3\ (+1.3)$ & $96.8\ (+1.8)$ & $7.83$ & $6.84\ (-12.7)$ & $6.42\ (-18.0)$ \\
$20$ & $10$ & $52.0$ & $30.0$ & $29.3$ & $93.1$ & $93.7\ (+0.7)$ & $94.4\ (+1.5)$ & $5.22$ & $4.62\ (-11.4)$ & $3.99\ (-23.5)$ & $94.6$ & $96.0\ (+1.5)$ & $96.7\ (+2.2)$ & $8.13$ & $7.00\ (-13.9)$ & $6.48\ (-20.3)$ \\
$21$ & $10$ & $57.2$ & $28.4$ & $27.4$ & $92.6$ & $93.5\ (+0.9)$ & $94.2\ (+1.7)$ & $5.56$ & $4.78\ (-14.1)$ & $4.17\ (-25.0)$ & $94.6$ & $96.5\ (+2.0)$ & $97.0\ (+2.5)$ & $9.39$ & $7.98\ (-15.1)$ & $7.60\ (-19.0)$ \\
$22$ & $10$ & $62.7$ & $27.0$ & $26.4$ & $92.2$ & $93.3\ (+1.2)$ & $93.9\ (+1.9)$ & $5.87$ & $4.90\ (-16.5)$ & $4.35\ (-25.9)$ & $93.8$ & $95.8\ (+2.2)$ & $96.1\ (+2.5)$ & $8.71$ & $7.08\ (-18.7)$ & $6.85\ (-21.4)$ \\
$23$ & $10$ & $68.4$ & $25.8$ & $24.9$ & $91.8$ & $93.1\ (+1.5)$ & $93.7\ (+2.1)$ & $6.20$ & $5.01\ (-19.2)$ & $4.50\ (-27.4)$ & $93.8$ & $96.4\ (+2.8)$ & $96.7\ (+3.1)$ & $9.94$ & $7.98\ (-19.7)$ & $7.72\ (-22.3)$ \\
$24$ & $10$ & $74.2$ & $24.6$ & $23.8$ & $91.4$ & $93.0\ (+1.7)$ & $93.5\ (+2.3)$ & $6.49$ & $5.11\ (-21.3)$ & $4.64\ (-28.5)$ & $93.4$ & $96.2\ (+3.0)$ & $96.4\ (+3.2)$ & $10.17$ & $8.06\ (-20.7)$ & $7.91\ (-22.2)$ \\
$25$ & $10$ & $80.2$ & $23.6$ & $22.8$ & $91.0$ & $92.9\ (+2.0)$ & $93.3\ (+2.5)$ & $6.80$ & $5.22\ (-23.3)$ & $4.81\ (-29.3)$ & $93.0$ & $96.2\ (+3.5)$ & $96.2\ (+3.4)$ & $10.44$ & $8.06\ (-22.8)$ & $8.09\ (-22.5)$ \\
$26$ & $10$ & $86.4$ & $22.6$ & $21.8$ & $90.7$ & $92.7\ (+2.3)$ & $93.2\ (+2.7)$ & $7.08$ & $5.29\ (-25.2)$ & $4.95\ (-30.1)$ & $92.6$ & $96.2\ (+3.8)$ & $96.2\ (+3.8)$ & $10.68$ & $8.07\ (-24.5)$ & $8.06\ (-24.5)$ \\
$27$ & $10$ & $92.8$ & $21.7$ & $21.0$ & $90.3$ & $92.7\ (+2.6)$ & $93.0\ (+3.0)$ & $7.36$ & $5.34\ (-27.5)$ & $5.05\ (-31.4)$ & $92.3$ & $96.1\ (+4.1)$ & $96.0\ (+4.0)$ & $10.92$ & $8.11\ (-25.7)$ & $8.17\ (-25.1)$ \\
$28$ & $10$ & $99.4$ & $20.9$ & $20.2$ & $90.0$ & $92.5\ (+2.8)$ & $92.9\ (+3.2)$ & $7.63$ & $5.42\ (-29.0)$ & $5.14\ (-32.7)$ & $91.9$ & $96.0\ (+4.4)$ & $95.9\ (+4.4)$ & $11.16$ & $8.12\ (-27.2)$ & $8.18\ (-26.7)$ \\
$29$ & $10$ & $106.2$ & $20.2$ & $19.3$ & $89.6$ & $92.4\ (+3.1)$ & $92.7\ (+3.4)$ & $7.88$ & $5.48\ (-30.5)$ & $5.24\ (-33.5)$ & $92.0$ & $96.4\ (+4.7)$ & $96.4\ (+4.7)$ & $12.22$ & $9.14\ (-25.2)$ & $9.11\ (-25.5)$ \\
$30$ & $10$ & $113.1$ & $19.5$ & $18.6$ & $89.3$ & $92.3\ (+3.4)$ & $92.6\ (+3.7)$ & $8.15$ & $5.53\ (-32.1)$ & $5.34\ (-34.4)$ & $91.7$ & $96.4\ (+5.1)$ & $96.4\ (+5.1)$ & $12.45$ & $9.08\ (-27.0)$ & $9.13\ (-26.7)$ \\
\hline
\hline
\end{tabular}
\end{center}
\end{table}
\end{landscape}
}

{\tabcolsep = 1.5mm
\begin{landscape}
\begin{table}
\caption{Same as Table~\ref{tab_eff1}, but here $R$ and $n$ are $0.9$ and $1.5$, respectively.}
\label{tab_eff2}
\begin{center}
\begin{tabular}{ccccccccccccccccccc}
\hline
\hline
&&&&& \multicolumn{3}{c}{\effmax} & \multicolumn{3}{c}{\effmaxbg} & \multicolumn{3}{c}{\effcross} & \multicolumn{3}{c}{\effcrossbg} \\ \cmidrule(lr){6-8} \cmidrule(lr){9-11} \cmidrule(lr){12-14} \cmidrule(lr){15-17}

\multicolumn{1}{c}{$\rho_1$} & \multicolumn{1}{c}{$\rho_2$} & \multicolumn{1}{c}{$L$} & \multicolumn{1}{c}{\Tmax} & \multicolumn{1}{c}{\Tcross} & Win. & Quad. & Cubic & Win. & Quad. & Cubic & Win. & Quad. & Cubic & Win. & Quad. & Cubic \\

\multicolumn{1}{c}{(mm)} & \multicolumn{1}{c}{(mm)} & \multicolumn{1}{c}{(mm)} & \multicolumn{1}{c}{($^\circ$)} & \multicolumn{1}{c}{($^\circ$)} & \multicolumn{1}{c}{(\%)} & \multicolumn{1}{c}{(\%)} & \multicolumn{1}{c}{(\%)} & \multicolumn{1}{c}{(\%)} & \multicolumn{1}{c}{(\%)} & \multicolumn{1}{c}{(\%)} & \multicolumn{1}{c}{(\%)} & \multicolumn{1}{c}{(\%)} & \multicolumn{1}{c}{(\%)} & \multicolumn{1}{c}{(\%)} & \multicolumn{1}{c}{(\%)} & \multicolumn{1}{c}{(\%)} \\ \hline
$18$ & $10$ & $41.9$ & $33.7$ & $32.2$ & $76.4$ & $76.7\ (+0.4)$ & $77.3\ (+1.1)$ & $3.26$ & $3.01\ (-7.8)$ & $2.37\ (-27.3)$ & $79.0$ & $79.8\ (+1.0)$ & $80.5\ (+2.0)$ & $6.92$ & $6.33\ (-8.6)$ & $5.64\ (-18.5)$ \\
$19$ & $10$ & $46.9$ & $31.8$ & $30.3$ & $75.8$ & $76.1\ (+0.5)$ & $76.8\ (+1.3)$ & $3.57$ & $3.18\ (-10.9)$ & $2.51\ (-29.6)$ & $78.3$ & $79.3\ (+1.3)$ & $80.0\ (+2.2)$ & $7.18$ & $6.37\ (-11.2)$ & $5.73\ (-20.3)$ \\
$20$ & $10$ & $52.0$ & $30.0$ & $28.7$ & $75.2$ & $75.6\ (+0.6)$ & $76.3\ (+1.5)$ & $3.86$ & $3.34\ (-13.5)$ & $2.61\ (-32.4)$ & $77.6$ & $79.0\ (+1.7)$ & $79.7\ (+2.7)$ & $7.42$ & $6.33\ (-14.6)$ & $5.71\ (-23.1)$ \\
$21$ & $10$ & $57.2$ & $28.4$ & $27.2$ & $74.6$ & $75.2\ (+0.8)$ & $75.8\ (+1.7)$ & $4.13$ & $3.51\ (-15.1)$ & $2.85\ (-31.0)$ & $77.0$ & $78.5\ (+1.9)$ & $79.2\ (+2.8)$ & $7.65$ & $6.46\ (-15.5)$ & $5.90\ (-22.8)$ \\
$22$ & $10$ & $62.7$ & $27.0$ & $25.8$ & $74.0$ & $74.7\ (+1.0)$ & $75.4\ (+1.9)$ & $4.36$ & $3.58\ (-17.8)$ & $2.95\ (-32.3)$ & $76.4$ & $78.1\ (+2.2)$ & $78.8\ (+3.1)$ & $7.84$ & $6.49\ (-17.2)$ & $5.97\ (-23.9)$ \\
$23$ & $10$ & $68.4$ & $25.8$ & $24.4$ & $73.5$ & $74.4\ (+1.2)$ & $75.0\ (+2.1)$ & $4.63$ & $3.71\ (-19.8)$ & $3.07\ (-33.6)$ & $76.3$ & $78.2\ (+2.6)$ & $78.8\ (+3.4)$ & $8.75$ & $7.26\ (-17.0)$ & $6.79\ (-22.4)$ \\
$24$ & $10$ & $74.2$ & $24.6$ & $23.3$ & $73.0$ & $74.0\ (+1.4)$ & $74.7\ (+2.3)$ & $4.85$ & $3.82\ (-21.3)$ & $3.14\ (-35.3)$ & $75.7$ & $77.8\ (+2.7)$ & $78.6\ (+3.7)$ & $8.94$ & $7.36\ (-17.6)$ & $6.75\ (-24.5)$ \\
$25$ & $10$ & $80.2$ & $23.6$ & $22.5$ & $72.4$ & $73.7\ (+1.7)$ & $74.3\ (+2.6)$ & $5.08$ & $3.85\ (-24.2)$ & $3.33\ (-34.5)$ & $74.7$ & $77.2\ (+3.3)$ & $77.6\ (+3.8)$ & $8.44$ & $6.56\ (-22.3)$ & $6.30\ (-25.3)$ \\
$26$ & $10$ & $86.4$ & $22.6$ & $21.4$ & $72.0$ & $73.4\ (+2.0)$ & $74.0\ (+2.8)$ & $5.30$ & $3.93\ (-25.9)$ & $3.41\ (-35.7)$ & $74.7$ & $77.3\ (+3.5)$ & $77.9\ (+4.4)$ & $9.28$ & $7.35\ (-20.8)$ & $6.90\ (-25.6)$ \\
$27$ & $10$ & $92.8$ & $21.7$ & $20.5$ & $71.5$ & $73.1\ (+2.3)$ & $73.7\ (+3.1)$ & $5.53$ & $3.98\ (-27.9)$ & $3.48\ (-37.1)$ & $74.1$ & $77.0\ (+3.9)$ & $77.6\ (+4.7)$ & $9.47$ & $7.35\ (-22.4)$ & $6.94\ (-26.7)$ \\
$28$ & $10$ & $99.4$ & $20.9$ & $20.0$ & $71.1$ & $72.8\ (+2.5)$ & $73.4\ (+3.3)$ & $5.73$ & $4.02\ (-29.9)$ & $3.60\ (-37.1)$ & $73.3$ & $76.4\ (+4.3)$ & $76.5\ (+4.4)$ & $9.00$ & $6.59\ (-26.7)$ & $6.56\ (-27.1)$ \\
$29$ & $10$ & $106.2$ & $20.2$ & $19.1$ & $70.6$ & $72.6\ (+2.7)$ & $73.1\ (+3.4)$ & $5.92$ & $4.04\ (-31.7)$ & $3.67\ (-38.0)$ & $73.2$ & $76.6\ (+4.7)$ & $76.8\ (+4.9)$ & $9.79$ & $7.26\ (-25.8)$ & $7.19\ (-26.5)$ \\
$30$ & $10$ & $113.1$ & $19.5$ & $18.4$ & $70.2$ & $72.3\ (+3.1)$ & $72.8\ (+3.7)$ & $6.11$ & $4.14\ (-32.3)$ & $3.70\ (-39.4)$ & $72.7$ & $76.3\ (+4.9)$ & $76.6\ (+5.4)$ & $9.92$ & $7.37\ (-25.7)$ & $7.10\ (-28.5)$ \\
\hline
\hline
\end{tabular}
\end{center}
\end{table}
\end{landscape}
}

A more realistic case in which $R$ and $n$ are assumed to be $0.9$ and $1.5$ is shown in Figures~\ref{fig_canCurve}(c) and (d). The optimized quadratic and cubic \Bezier\ cones again outperform the normal Winston cone, exhibiting higher collection efficiencies for signal photons and lower efficiencies for stray light. The values of \effmax, \effmaxbg, \effcross, and \effcrossbg\ for the realistic case are presented in Table~\ref{tab_eff2}. When $\rho_1=20$~mm, \effcross\ for the Winston and optimized cubic \Bezier\ cones are $77.6$\% and $79.7$\%, respectively, and \effcrossbg\ are $7.42$\% and $5.71$\%, respectively. Therefore, we can gain a $2.7$\% higher collection efficiency for signal photons and a $23.1$\% lower efficiency for the background in this case.

Figure~\ref{fig_canGain} shows \effcross\ vs. \Tcross, and \effcrossbg\ vs. \Tcross\ for the ideal ($R=1.0$ and $n=1.0$) and realistic ($R=0.9$ and $n=1.5$) cases. Better \effcross\ and \effcrossbg\ values are achieved by the optimized \Bezier\ curves.

\begin{figure}
  \begin{center}
    \includegraphics[width=9cm]{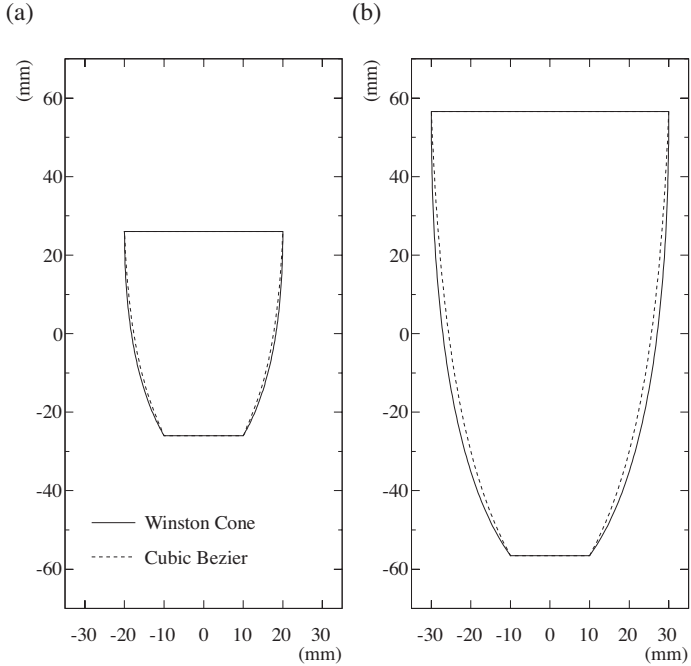}
   \caption{(a) Vertical sections of a hexagonal Winston cone (solid) and the optimized cone with a cubic \Bezier\ curve (dashed). Here, $\rho_1$, $\rho_2$, $R$, and $n$ are $10$~mm, $20$~mm, $1.0$, and $1.0$, respectively. The quadratic \Bezier\ and $R=0.9$/$n=1.5$ cases are not shown because the difference is indistinguishable by eye. (b) Same as (a) but $\rho_1$ and $\rho_2$ are $30$~mm and $20$~mm, respectively.}
    \label{fig_canShape}
  \end{center}
\end{figure}

Two of the optimized shapes listed in Table~\ref{tab_coords} are drawn in Figure~\ref{fig_canShape}. The optimized curves are slightly narrower than Winston's paraboloids. The widths at the middle are $1.2$ mm smaller for $\rho_1=20$~mm and $2.9$~mm smaller for $\rho_1=30$~mm.

\section{Conclusion}

We simulated the collection efficiency of hexagonal light collectors with different $\rho_1$, $R$, and $n$ values. Using quadratic or cubic \Bezier\ curves instead of the traditional Winston cones, we found that a few percent higher collection efficiencies for signal photons and a few tens of percent lower efficiencies for stray background light could be simultaneously achieved. In this way, we can improve the signal-to-noise ratio of atmospheric Cherenkov photons induced by very-high-energy gamma rays against the night-sky background. Thus, the gamma-ray detection efficiency for future projects such as the Cherenkov Telescope Array (CTA) \cite{Actis:2011:Design-concepts-for-the-Cherenkov-Telesc} can be improved without any additional cost or new technology. This improvement is expected to yield lower energy thresholds, larger effective areas, and higher energy and angular resolutions of very-high-energy gamma rays.

\section*{Acknowledgment}
We thank Dr. Masaaki Hayashida and Dr. Takayuki Saito for helpful discussions. The author is supported by a Grant-in-Aid for JSPS Fellows. A part of this work was performed using comuter resources at Institicte for Cosmic Ray Research (ICRR), the University of Tokyo.




\bibliographystyle{model1-num-names}
\bibliography{OkumuraCone}







\end{document}